\documentclass[aps,prl,twocolumn,superscriptaddress,showpacs,epsfig]{revtex4}
\usepackage{graphicx}
\usepackage{graphics}
\usepackage{epsfig}
\usepackage{subfigure}
\usepackage{longtable}
\newcommand{\met} {\hbox{$E$\kern-0.55em\lower-0.15ex\hbox{/}}_T}
\newcommand{\vetmet} {\hbox{$\vec{E}$\kern-0.55em\lower-0.15ex\hbox{/}}_T}
\begin{document}
\title{Measurement of the $t\bar t$ Production Cross Section in  $p\bar p$ 
Collisions at $\sqrt{s} = 1.96$ TeV using Missing $E_T$+jets
Events with Secondary Vertex $b$-Tagging}

\affiliation{Institute of Physics, Academia Sinica, Taipei, Taiwan 11529, Republic of China} 
\affiliation{Argonne National Laboratory, Argonne, Illinois 60439} 
\affiliation{Institut de Fisica d'Altes Energies, Universitat Autonoma de Barcelona, E-08193, Bellaterra (Barcelona), Spain} 
\affiliation{Baylor University, Waco, Texas  76798} 
\affiliation{Istituto Nazionale di Fisica Nucleare, University of Bologna, I-40127 Bologna, Italy} 
\affiliation{Brandeis University, Waltham, Massachusetts 02254} 
\affiliation{University of California, Davis, Davis, California  95616} 
\affiliation{University of California, Los Angeles, Los Angeles, California  90024} 
\affiliation{University of California, San Diego, La Jolla, California  92093} 
\affiliation{University of California, Santa Barbara, Santa Barbara, California 93106} 
\affiliation{Instituto de Fisica de Cantabria, CSIC-University of Cantabria, 39005 Santander, Spain} 
\affiliation{Carnegie Mellon University, Pittsburgh, PA  15213} 
\affiliation{Enrico Fermi Institute, University of Chicago, Chicago, Illinois 60637} 
\affiliation{Joint Institute for Nuclear Research, RU-141980 Dubna, Russia} 
\affiliation{Duke University, Durham, North Carolina  27708} 
\affiliation{Fermi National Accelerator Laboratory, Batavia, Illinois 60510} 
\affiliation{University of Florida, Gainesville, Florida  32611} 
\affiliation{Laboratori Nazionali di Frascati, Istituto Nazionale di Fisica Nucleare, I-00044 Frascati, Italy} 
\affiliation{University of Geneva, CH-1211 Geneva 4, Switzerland} 
\affiliation{Glasgow University, Glasgow G12 8QQ, United Kingdom} 
\affiliation{Harvard University, Cambridge, Massachusetts 02138} 
\affiliation{Division of High Energy Physics, Department of Physics, University of Helsinki and Helsinki Institute of Physics, FIN-00014, Helsinki, Finland} 
\affiliation{University of Illinois, Urbana, Illinois 61801} 
\affiliation{The Johns Hopkins University, Baltimore, Maryland 21218} 
\affiliation{Institut f\"{u}r Experimentelle Kernphysik, Universit\"{a}t Karlsruhe, 76128 Karlsruhe, Germany} 
\affiliation{High Energy Accelerator Research Organization (KEK), Tsukuba, Ibaraki 305, Japan} 
\affiliation{Center for High Energy Physics: Kyungpook National University, Taegu 702-701; Seoul National University, Seoul 151-742; and SungKyunKwan University, Suwon 440-746; Korea} 
\affiliation{Ernest Orlando Lawrence Berkeley National Laboratory, Berkeley, California 94720} 
\affiliation{University of Liverpool, Liverpool L69 7ZE, United Kingdom} 
\affiliation{University College London, London WC1E 6BT, United Kingdom} 
\affiliation{Massachusetts Institute of Technology, Cambridge, Massachusetts  02139} 
\affiliation{Institute of Particle Physics: McGill University, Montr\'{e}al, Canada H3A~2T8; and University of Toronto, Toronto, Canada M5S~1A7} 
\affiliation{University of Michigan, Ann Arbor, Michigan 48109} 
\affiliation{Michigan State University, East Lansing, Michigan  48824} 
\affiliation{Institution for Theoretical and Experimental Physics, ITEP, Moscow 117259, Russia} 
\affiliation{University of New Mexico, Albuquerque, New Mexico 87131} 
\affiliation{Northwestern University, Evanston, Illinois  60208} 
\affiliation{The Ohio State University, Columbus, Ohio  43210} 
\affiliation{Okayama University, Okayama 700-8530, Japan} 
\affiliation{Osaka City University, Osaka 588, Japan} 
\affiliation{University of Oxford, Oxford OX1 3RH, United Kingdom} 
\affiliation{University of Padova, Istituto Nazionale di Fisica Nucleare, Sezione di Padova-Trento, I-35131 Padova, Italy} 
\affiliation{University of Pennsylvania, Philadelphia, Pennsylvania 19104} 
\affiliation{Istituto Nazionale di Fisica Nucleare Pisa, Universities of Pisa, Siena and Scuola Normale Superiore, I-56127 Pisa, Italy} 
\affiliation{University of Pittsburgh, Pittsburgh, Pennsylvania 15260} 
\affiliation{Purdue University, West Lafayette, Indiana 47907} 
\affiliation{University of Rochester, Rochester, New York 14627} 
\affiliation{The Rockefeller University, New York, New York 10021} 
\affiliation{Istituto Nazionale di Fisica Nucleare, Sezione di Roma 1, University of Rome ``La Sapienza," I-00185 Roma, Italy} 
\affiliation{Rutgers University, Piscataway, New Jersey 08855} 
\affiliation{Texas A\&M University, College Station, Texas 77843} 
\affiliation{Istituto Nazionale di Fisica Nucleare, University of Trieste/\ Udine, Italy} 
\affiliation{University of Tsukuba, Tsukuba, Ibaraki 305, Japan} 
\affiliation{Tufts University, Medford, Massachusetts 02155} 
\affiliation{Waseda University, Tokyo 169, Japan} 
\affiliation{Wayne State University, Detroit, Michigan  48201} 
\affiliation{University of Wisconsin, Madison, Wisconsin 53706} 
\affiliation{Yale University, New Haven, Connecticut 06520} 
\author{A.~Abulencia}
\affiliation{University of Illinois, Urbana, Illinois 61801}
\author{D.~Acosta}
\affiliation{University of Florida, Gainesville, Florida  32611}
\author{J.~Adelman}
\affiliation{Enrico Fermi Institute, University of Chicago, Chicago, Illinois 60637}
\author{T.~Affolder}
\affiliation{University of California, Santa Barbara, Santa Barbara, California 93106}
\author{T.~Akimoto}
\affiliation{University of Tsukuba, Tsukuba, Ibaraki 305, Japan}
\author{M.G.~Albrow}
\affiliation{Fermi National Accelerator Laboratory, Batavia, Illinois 60510}
\author{D.~Ambrose}
\affiliation{Fermi National Accelerator Laboratory, Batavia, Illinois 60510}
\author{S.~Amerio}
\affiliation{University of Padova, Istituto Nazionale di Fisica Nucleare, Sezione di Padova-Trento, I-35131 Padova, Italy}
\author{D.~Amidei}
\affiliation{University of Michigan, Ann Arbor, Michigan 48109}
\author{A.~Anastassov}
\affiliation{Rutgers University, Piscataway, New Jersey 08855}
% ...
\collaboration{CDF Collaboration}
\noaffiliation



\begin{abstract}
\noindent
We present a measurement of the $t\bar t$ production cross section
in $p\bar p$ collisions at $\sqrt{s}=1.96$ TeV which uses 
events with an inclusive signature of significant missing transverse
energy and jets. This is the first measurement which makes no 
explicit lepton identification requirements, so that sensitivity to 
$W\to\tau\nu$ decays is maintained. Heavy flavor jets from top quark decay are 
identified with a secondary vertex tagging algorithm. From $311$ pb$^{-1}$ of 
data collected by the Collider Detector at Fermilab we measure a production 
cross section of $5.8 \pm 1.2$(stat.)$^{+0.9}_{-0.7}$(syst.) pb for a top 
quark mass of $178$ GeV/$c^2$, in agreement with previous determinations and 
standard model predictions.
\end{abstract}

\maketitle

%
%
\noindent
At the Tevatron $p\bar p$ collider top quarks are produced mainly in pairs
through quark-antiquark annihilation and gluon-gluon fusion processes.
In the standard model (SM) the calculated cross section for pair production 
is $6.1^{+0.6}_{-0.8}$ pb \cite{topthxsec} for a top mass of $178$ GeV/$c^2$
\cite{topmass}, 
and varies linearly with a slope of $-0.2$ pb/GeV/$c^2$ with the top quark 
mass in the range $170$ GeV/$c^2 < m_t < 190$ GeV/$c^2$. 
Because the Cabibbo-Kobayashi-Maskawa matrix element $V_{tb}$ is close
to unity and $m_t$ is large, the SM top quark decays  to a $W$ boson 
and a $b$ quark almost $100\%$ of the time. The final state of top quark pair 
production thus includes two $W$ bosons and two $b$-quark jets. When only one 
$W$ decays leptonically, the $t\bar t$ event typically contains a charged 
lepton, missing transverse energy from the undetected neutrino, and four high 
transverse energy jets, two of which originate from $b$ quarks
\cite{CDFcoord}.
Previous cross section analyses \cite{cdfxsec1,cdfxsec2secvtx,d0all} select
this distinct $t\bar t$ signature by requiring  well-identified leptons 
($e,~\mu$) with high transverse momentum. 
In this Letter we describe a $t\bar t$ production cross section measurement
which is sensitive to leptonic $W$ decays regardless of the lepton type, and
has a sizable acceptance to $\tau$ decays of the $W$ boson. 
The direct identification of $\tau$ from $W\to \tau\nu$ decays suffers from a very low
efficiency, thus our  measurement, using data collected by a multijet trigger, selects 
top decays by inclusively requiring a high-$p_T$ neutrino signature rather 
than charged lepton identification. Events with well-identified high-$p_T$  
electrons or muons are explicitly vetoed in order to avoid statistical overlap 
and provide complementary results with respect to lepton-based measurements.
%
%
%

Results reported in this Letter  are obtained from $311$ pb$^{-1}$ of data
from $p\bar p$ collisions at $\sqrt{s} =1.96$ TeV recorded by the Collider 
Detector at Fermilab (CDF\,II).
The CDF\,II detector has been described in detail elsewhere 
\cite{CDFII}. It consists of a magnetic spectrometer surrounded 
by calorimeter systems  and muon chambers. The momenta of charged particles 
are measured up to a pseudorapidity of $|\eta|=1.0$ in the central tracking 
chamber, which is inside a 1.4 T superconducting solenoidal magnet. 
Microstrip silicon vertex detectors, located immediately outside the beampipe,
provide precise track reconstruction useful for vertexing and 
extend the $|\eta|$ coverage of the tracking system up to $|\eta|=2.0$.
Electromagnetic and hadronic sampling calorimeters, arranged in a 
projective-tower geometry, surround the tracking systems and measure the 
energy and direction of electrons, photons, and jets, providing good 
hermeticity in the pseudorapidity range $|\eta|<3.6$. In addition, 
the calorimeter system allows the detection of high-$p_T$ neutrinos by 
the measurement of the missing transverse energy, $\vetmet$ \cite{CDFcoord}.
Muon systems reside outside the calorimeters and consist of layers of
drift chambers which allow the reconstruction of track segments for 
penetrating particles.
The beam luminosity is determined using gas Cherenkov counters (CLC)
located in the region $3.7 < |\eta|<4.7$ which measure the average number of 
inelastic $p\bar p$ collisions per bunch crossing. The uncertainty on the 
luminosity measurement, due to CLC acceptance and theoretical
uncertainties on the total inelastic $p\bar p$ cross section, is 6\%
\cite{CLC}.
%
%
%

The data sample used in this analysis is collected by a multijet trigger 
which requires four or more  $E_T\ge 15$ GeV clusters  of contiguous 
calorimeter towers, and a total transverse 
energy clustered in the calorimeter of $\sum E_T \ge 125$ GeV. The 
initial data sample 
consists of $4.2$ million events.
The understanding of signal acceptances and efficiencies relies on a detailed 
simulation of the production processes and  the detector response. 
Inclusively decaying $t\bar t$ events,  assuming a top quark mass of $178$ 
GeV/$c^2$, are simulated using {\scshape{pythia}} v6.2 \cite{pythia} and 
{\scshape{herwig}} v6.5 \cite{herwig} generators in conjunction with the 
{\scshape{cteq5l}} \cite{CTEQ} parton distribution functions, 
{\scshape{qq}}v9.1 \cite{QQ} for the modeling of $b$ and $c$ hadron decays, 
and a full simulation of the CDF\,II detector.
Jets are identified as groups of calorimeter tower energy deposits which 
fall within a cone of radius 
$\Delta R = \sqrt{\Delta \phi^2 + \Delta\eta^2} \le 0.4$. Jet energies,
after the absolute energy scale setting,  
are corrected for calorimeter non-linearity, losses in the gap between 
towers, and multiple interactions \cite{JetCorr}. 

The $t\bar t$ signature used in the present study ($\met$+jets) consists of
large missing transverse energy, $\vetmet$, associated with the 
neutrino from the leptonic decay of a $W$ boson, and jets. 
Since  the $\met$ resolution, $\sigma(\met)$, 
is observed to degrade as a function of the total transverse energy of the 
event, in the form $\sigma(\met) \propto \sqrt{\sum E_T}$ \cite{topevid},
the missing $E_T$ significance, defined by 
$\met^{sig} = |\vec{\met}| / \sqrt{\sum E_T}$, is used for event selection.
Besides the requirement of large $\met^{sig}$, further background rejection can
be achieved by exploiting $\vetmet$-related geometrical properties.
The neutrino direction for $t\bar t \to \met$ + jets events is in general 
uncorrelated with the jet directions in the transverse plane. On the contrary, 
background events for which the $\vetmet$ is mainly due to jet 
energy mis-measurement or to $b$-quark semileptonic decays are more 
likely to have $\vetmet$ aligned with a jet direction. The minimum 
$\Delta\phi$ difference between the $\vetmet$ and any jet 
in the event, $\min\Delta\phi(\met,jets)$, is used as an analysis cut.
Events containing identified high-$p_T$  electrons or muons, 
as defined in \cite{cdfxsec2secvtx}, are removed, in order to increase the
relative contribution of $W\to \tau \nu$ decays and to provide a 
statistically independent sample with respect to other lepton+jets cross 
section analyses. 

To reject events with only light quark or gluon jets, we require at least one 
jet to be identified as originating from a $b$-quark.
The presence of $b$-jets (``tags'') is established by the identification of
secondary decay vertices using the {\scshape{secvtx}} 
algorithm \cite{cdfxsec2secvtx}.
%
%
An optimization procedure, aimed at minimizing the relative statistical
uncertainty on the cross section measurement, 
sets $N_{jet}(E_T\ge 15$ GeV; $|\eta|\le 2.0)\ge 4$, 
$\met^{sig}\ge 4.0$ GeV $^{1/2}$,  and $\min\Delta\phi(\met,jets)\ge 0.4$ rad 
as the best kinematical selection cuts. After these requirements the
data sample is reduced to $597$ events with an expected pre-tag
signal to background 
ratio $S/B\sim 1/5$.  
%
%
%
%
\begin{figure}[Hb!]
\begin{center}
\epsfig{figure=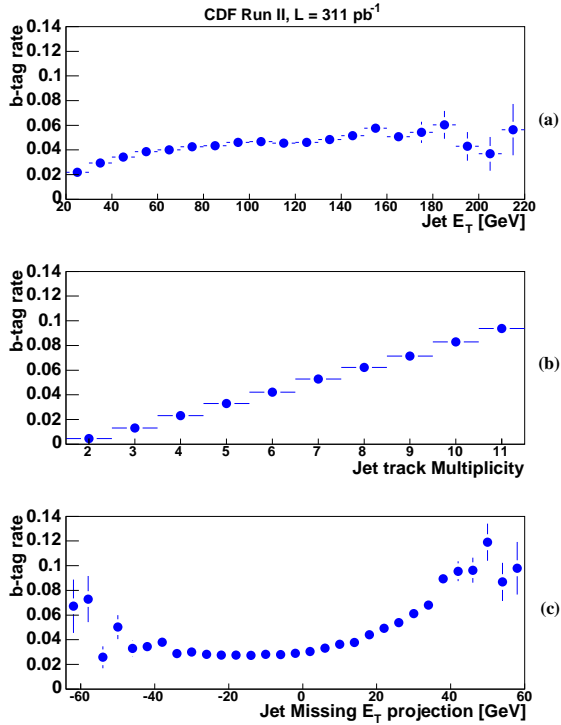,
	width=.9\linewidth, clip=}
\caption{Tagging rate probability parametrization in the multijet sample 
         as a function of jet $E_T$ (a), track multiplicity (b), 
         and missing $E_T$ projection (c).}
\label{f:tagrates}
\end{center}
\end{figure}
%
%
%

\begin{figure}[Ht!]
\begin{center}

{\epsfig{figure=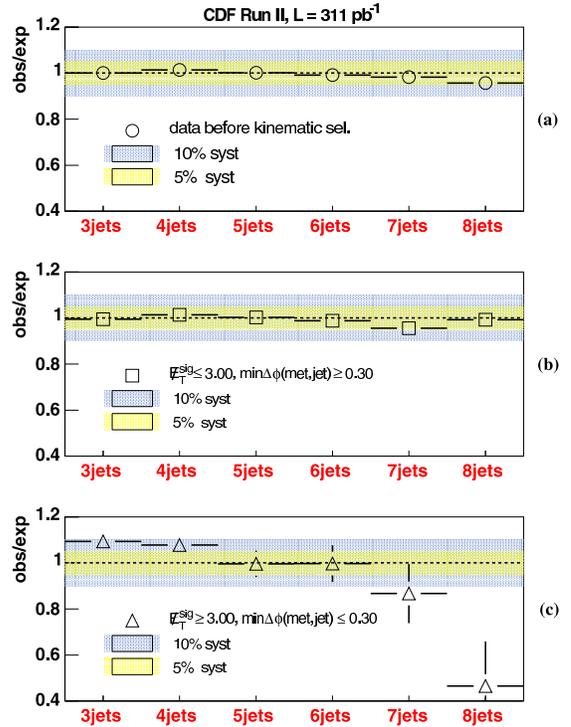,
                     width=.9\linewidth,clip=}}
\caption{Ratio of observed to expected $b$-tags as a function of the 
         number of jets in the data before the optimized kinematical 
         selection on $\met^{sig}$ and $\min\Delta\phi(\met,jets)$ (a), and 
         in control samples: $\met^{sig} \le 3.0$ GeV$^{1/2}$, 
         $\min\Delta\phi(\met,jets)\ge 0.3$ rad (b); and 
         $\met^{sig} \ge 3.0$ GeV$^{1/2}$, $\min\Delta\phi(\met,jets)\le 
         0.3$ rad (c). }
\label{f:tagratescheck}
\end{center}
\end{figure}

Background events with $b$-tags arise from QCD heavy flavor production, 
electroweak production of $W$ bosons associated with heavy flavor jets, 
and from false identification by the {\scshape{secvtx}} algorithm.
The overall number of background $b$-tags is estimated from the multijet 
sample by applying a parametrization of the per-jet tagging probability. 
The tagging probability is calculated using $\sim 879,000$ multijet data 
events with exactly three jets having $E_T\ge 15$ GeV and $|\eta|\le 2.0$. 
It is parametrized on a per-jet basis as a function of the jet $E_T$, track 
multiplicity, and the projection of the $\vetmet$ along the jet direction, 
defined by $\met^{prj} = \vetmet \cdot\cos\Delta\phi(\met,jet)$.  
The $t\bar t$ contamination in this control sample is negligible.
The jet $b$-tagging rate, calculated as the ratio of the number of 
$b$-tagged jets to the number of jets with at least two good tracks with 
hits in the silicon detector, is shown in Fig.~\ref{f:tagrates}. 
The $\met^{prj}$ parametrization accounts for background $b$-tags from
$b\bar b$ production processes, whose $b$-tag rate is enhanced at high 
positive values of the $\met^{prj}$ due to the correlation of $\vetmet$ and 
$b$-jet direction in the case of a semileptonic $b$-quark decay.
The extrapolation of the $3$-jet $b$-tag rate to higher jet multiplicity 
events is checked by comparing the predicted and observed $b$-tag rates as a 
function of the number of jets in the multijet sample without the kinematical 
selection on $\met^{sig}$ and $\min\Delta\phi(\met,jets)$, 
as shown in Fig.~\ref{f:tagratescheck}(a). 
The capability of the parametrization to track possible sample 
composition changes introduced by the kinematical selection is established by 
constructing the ratio of observed over expected $b$-tags in two 
separate control samples of multijet data, depleted of signal contaminations, as shown in 
Fig.~\ref{f:tagratescheck}(b) and Fig.~\ref{f:tagratescheck}(c), respectively.
The first control sample is defined by 
$\met^{sig} \le 3.0$ GeV$^{1/2}$, $\min\Delta\phi(\met,jets)\ge 
0.3$ rad; the second as $\met^{sig} \ge 3.0$ GeV$^{1/2}$, 
$\min\Delta\phi(\met,jets)\le 0.3$ rad. 
The $b$-tag rate parametrization is found to predict the 
non-$t\bar t$ background to within $10\%$ in the $4\le N_{jet} \le 6$
region where $96.4\%$ of the $t\bar t$ signal is expected.  This value is 
thus adopted as the systematic uncertainty on the background prediction. 
Additional checks, using low jet multiplicity events from leptonic
$W$ data samples,  show agreement between observed 
and predicted $b$-tags to within the quoted uncertainty.
%
%
%
%
%
%
%
\begin{figure}[Ht!]
\begin{center}
\epsfig{figure=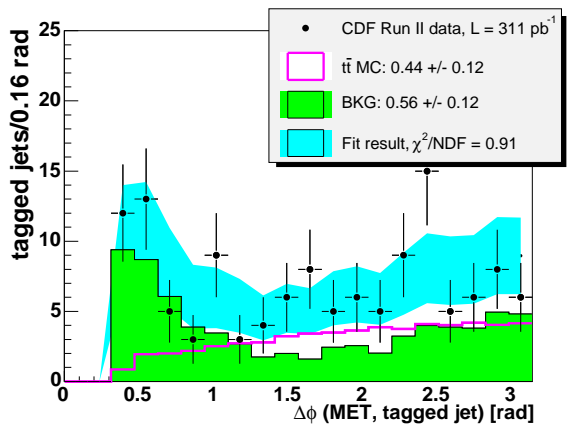,
                     width=.9\linewidth,clip=}
\caption{$\Delta\phi(\met,tagged~jet)$ distribution for data after
         kinematical selection and with at least one $b$-tagged jet.
         The data are fit to the sum of $t\bar t$ signal and 
         background templates as described in the text. }
\label{f:fits1}
\end{center}
\end{figure}
%
%
%
\begin{table*}
 \begin{ruledtabular}
 \caption{ Number of $b$-tagged jets  expected from Monte Carlo $t\bar t$ production using 
           $\sigma(t\bar t)= 6.1$ pb ($m_{t} = 178$ GeV/$c^2$), 
           predicted by the tagging rate parametrization, and observed in the 
           selected sample, as a function of the jet multiplicity. The total
           uncertainty on the number of predicted background $b$-tags 
           is the sum in quadrature of the statistical uncertainty and of
           a $10\%$ systematic uncertainty. The number of background $b$-tags
           corrected for the $t\bar t$ contamination in the pre-tagging data 
           sample is also provided for the signal region.  
           Uncertainties on signal contributions
           are statistical only.}
\label{t:kintag}
\begin{tabular}{lrrrr}
 Number of jets & 3 & 4 & 5 & $\ge 6$ \\
 \hline
   $t\bar t\to ee$         & $0.08\pm 0.01$  & $  0.41\pm 0.03$  & $ 0.18\pm 0.02$  & $0.04\pm 0.01$ \\
   $t\bar t\to e\mu$       & $0.06\pm 0.01$  & $  0.29\pm 0.02$  & $ 0.11\pm 0.01$  & $0.05\pm 0.01$ \\
   $t\bar t\to \mu\mu$     & $0.01\pm 0.01$  & $  0.05\pm 0.01$  & $ 0.01\pm 0.01$  & $0.01\pm 0.01$ \\
   $t\bar t\to e\tau$      & $0.11\pm 0.01$  & $  0.93\pm 0.04$  & $ 0.38\pm 0.03$  & $0.15\pm 0.02$ \\
   $t\bar t\to \mu\tau$    & $0.05\pm 0.01$  & $  0.29\pm 0.02$  & $ 0.15\pm 0.02$  & $0.06\pm 0.01$ \\
   $t\bar t\to \tau\tau$   & $0.06\pm 0.01$  & $  0.58\pm 0.03$  & $ 0.26\pm 0.02$  & $0.05\pm 0.01$ \\
   $t\bar t\to e+jets$     & $0.68\pm 0.04$  & $  6.61\pm 0.11$  & $ 8.70\pm 0.13$  & $4.25\pm 0.09$ \\
   $t\bar t\to \mu+jets$   & $1.07\pm 0.04$  & $ 11.92\pm 0.15$  & $ 6.56\pm 0.11$  & $2.47\pm 0.07$ \\
   $t\bar t\to \tau+jets$  & $1.00\pm 0.04$  & $ 10.98\pm 0.14$  & $11.71\pm 0.15$  & $5.53\pm 0.10$ \\
   $t\bar t\to jets$       & $0.01\pm 0.01$  & $  0.09\pm 0.01$  & $ 0.14\pm 0.02$  & $0.22\pm 0.02$ \\
 \hline
   $t\bar t\to X$          & $3.13 \pm 0.08$  & $32.15 \pm 0.24$  & $28.14\pm 0.23$  & $12.83\pm 0.15$ \\

  Background 
  $b$-tagged jets   & $32.68\pm 3.46$ & $37.53 \pm 4.14$ & 
                             $ 21.44 \pm 2.76$ & $ 8.47 \pm 1.40$ \\
  Corrected background 
  $b$-tagged jets   & $-$ & $33.14 \pm 4.01$ & 
                             $ 17.58 \pm 2.85$ & $ 6.71 \pm 2.78$ \\

 Observed 
 $b$-tagged jets     & 31  & 53  & 55  &19 \\
 \end{tabular}
 \end{ruledtabular}
 \end{table*}

%
%
%

The sample selected with the optimized kinematical requirements described 
above and the requirement of at least one  $b$-tagged jet consists of $106$ 
events, containing a total of $127$ $b$-tagged jets. The number of $b$-tagged 
jets yielded by background processes in that sample is expected to be 
$N_{exp} = 67.4\pm 2.7\pm 6.7$. The first uncertainty contribution
is due to the limited number of events in the $3$-jet sample used for the 
tagging rate parametrization, while the second contribution is the $10\%$
systematic uncertainty on the $b$-tag rate parameterization discussed above
(Fig.~\ref{f:tagratescheck}). 
The $597$ pre-tag data events are expected to contain a non-negligible $t\bar t$ 
component, due to the $S/N$ enhancement provided by the kinematical selection, 
which yields a background overestimate. The $N_{exp}$ value is corrected, 
to $N_{exp}^\prime = 57.4\pm 8.1$, to account for this effect.
Table \ref{t:kintag} summarizes the number of $b$-tagged jets expected from
Monte Carlo simulation for each $t\bar t$ decay mode satisfying the kinematical 
and $\ge 1$ $b$-tag requirements, as well as the predicted and observed $b$-tags 
in the selected data sample as a function of the jet multiplicity of the events. In 
Table~\ref{t:kintag}, the numbers of $b$-tags from events with only three jets
are provided as a cross-check of the background. The excess 
in the number of $b$-tags, for $N(jet)\ge 4$, is ascribed to top pair 
production. Final states with $W\to \tau\nu$ decays account for 
$\sim 44\%$ of the signal acceptance. 

In order to further establish the $t\bar t$ signal in the selected data 
we perform binned likelihood fits to kinematical distributions. The stability 
of the fitting technique is checked using simulations with known signal 
fractions. We fit $\met$ and $\Delta\phi(\met,tagged~jet)$ data distributions 
to the sum of a $t\bar t$ and a background template. The former is obtained 
from Monte Carlo $t\bar t$ inclusive events; the latter is derived from the 
tagging rate parametrization applied to the data. Results for the 
$\Delta\phi(\met,tagged~jet)$ fit are displayed in 
Fig.~\ref{f:fits1}.  The fitted $t\bar t$ 
components in the selected data ($68\pm 12\%$ and $44 \pm 12\%$ for the 
$\met$ and $\Delta\phi(\met,tagged~jet)$ fits respectively) are in agreement 
with the overall prediction calculated from $b$-tag counting, before
any correction to account for the $t\bar t$ contamination in the pre-tagging
data sample
($47 \pm 5\%$ determined comparing the number of expected and observed
$b$-tags, for $N_{jet}\ge 4$, in Table~\ref{t:kintag}). 
%
%
%
\begin{table}
\begin{center}
\begin{ruledtabular}
\caption{Relevant sources of systematic uncertainty.}
\label{t:systsall}
\begin{tabular}{lr}
Source & relative error \\
\hline
\multicolumn{2}{c}{$\epsilon_{kin}$ systematics}\\

Generator dependence &   8.2  \% \\
Trigger acceptance   &   2.0 \% \\
ISR/FSR              &   2.0 \% \\
PDFs                 &   1.6 \% \\
Jet Energy Scale     &   1.5 \%  \\
\hline
\multicolumn{2}{c}{Others} \\                 					 
Background prediction method & 10.0 \%    \\
Luminosity measurement & 6.0 \%    \\
$\epsilon_{tag}^{ave}$ ({\scshape{secvtx}} scale factor)  &  5.8 \%   \\
\end{tabular}
\end{ruledtabular}
\end{center}

\end{table}
%
%
%

The efficiency of the trigger, the kinematical selection, and the $b$-tagging
algorithm, are evaluated using inclusive $t\bar t$ Monte Carlo events. The 
combined efficiency of the trigger and kinematical selection amounts to 
$\epsilon_{kin} = 4.88\% \pm 0.43 \%$ for a top mass of $m_t=178$ GeV/$c^2$, 
where the dominant uncertainty is determined by the comparison
of the results from the {\scshape{pythia}} and {\scshape{herwig}} generators.
%
Other 
sources of systematic uncertainty are evaluated by varying Monte Carlo 
generation settings
with respect to the default values, and by applying systematic variations of
the jet correction factors; the trigger acceptance 
uncertainty is determined by comparing trigger turn-on curves between Monte 
Carlo and data events (Table~\ref{t:systsall}).
The average number of $b$-tags per Monte Carlo $t\bar t$ event is found to be 
$\epsilon_{tag}^{ave}= 0.789\pm 0.046$, and it is corrected according to 
the Monte Carlo {\scshape{secvtx}} efficiency scale factor of $0.909\pm 0.060$ 
to reproduce the data $b$-tagging efficiency \cite{cdfxsec2secvtx}.
%
  
The cross section is calculated with a Poisson likelihood function
in which the maximum likelihood solution for $\sigma(t\bar t)$ is given by:
$\sigma(t\bar t) = \frac{N_{obs} - N_{exp}^\prime}
                       {\epsilon_{kin}\cdot \epsilon_{tag}^{ave}\cdot \mathcal{L}_{int}}$,
where $N_{obs}$ and $N_{exp}^{\prime}$ are the number of $b$-tagged 
jets observed and expected from background events by the tagging rate 
parametrization in the selected data, and ${\mathcal{L}}_{int}$ is the 
integrated luminosity of the multijet data sample. The input parameters: 
${\mathcal{L}}_{int}$, $\epsilon_{kin}$, $\epsilon_{tag}^{ave}$, and $N_{exp}^\prime$ 
are subject to Gaussian constraints.
%
%
%
%
With these input values we measure a top pair 
production cross section of $5.8\pm 1.2$(stat.)$^{+0.9}_{-0.7}$(syst.) pb.
%
Additional samples of inclusive $t\bar t$ Monte Carlo events generated with
different $m_t$ values in the range $[130, 230]$ GeV/$c^2$ are used to
compute the cross section measurement dependence on $m_t$. The cross section 
measurement changes by $\pm 0.05$ pb  for each $\mp 1$ GeV/$c^2$ change in the 
top quark mass from the initial value of $178$ GeV/$c^2$. 
For instance, we measure
$\sigma(t\bar t) = 6.0\pm 1.2$(stat.)$^{+0.9}_{-0.7}$(syst.) pb assuming a top
quark mass of $ 175$ GeV/$c^2$. The change is due to the varying signal
selection efficiency with top quark mass.
%
%

In conclusion we report the first measurement of the top pair production
cross section of $\sigma(t\bar t) = 5.8^{+1.5}_{-1.4}$ pb using 
an inclusive selection of $\met$+jets $t\bar t$ decays. The result
is complementary to other cross section analyses
\cite{cdfxsec1,cdfxsec2secvtx,d0all}, 
maintains high sensitivity with respect to $W\to\tau\nu$ $t\bar t$ decays, 
and is in good agreement with SM calculations 
\cite{topthxsec} and previous measurements.
%
%
%

We thank the Fermilab staff and the technical staffs of the participating 
institutions for their vital contributions. This work was supported by the 
U.S. Department of Energy and National Science Foundation; the Italian 
Istituto Nazionale di Fisica Nucleare; the Ministry of Education, Culture, 
Sports, Science and Technology of Japan; the Natural Sciences and Engineering 
Research Council of Canada; the National Science Council of the Republic of 
China; the Swiss National Science Foundation; the A.P. Sloan Foundation; the 
Bundesministerium f\"ur Bildung und Forschung, Germany; the Korean Science 
and Engineering Foundation and the Korean Research Foundation; the Particle 
Physics and Astronomy Research Council and the Royal Society, UK; the Russian 
Foundation for Basic Research; the Comisi\'on Interministerial de Ciencia y 
Tecnolog\'{\i}a, Spain; in part by the European Community's Human Potential 
Programme under contract HPRN-CT-2002-00292; and by the Academy of Finland. 

%
%

%

\end{document}